\documentclass{elsart}

\begin{document}

\begin{frontmatter}
\title{Solutions to the Quasi-flatness and Quasi-lambda Problems}
\author{John D. Barrow$^1$ and Jo\~ao Magueijo$^2$}
\address{$^1$Astronomy Centre, University of Sussex, Brighton BN1 9QJ, U.K.\\
$^2$Theoretical Physics, The Blackett Laboratory, \\
Imperial College, Prince Consort Rd., London, SW7 2BZ, U.K.}
\maketitle
\date{}

\begin{abstract}
Big Bang models of the Universe predict rapid domination by
curvature, a paradox known as the flatness problem. Solutions 
to this problem usually leave the Universe exactly flat for every
practical purpose. Explaining a nearly but not exactly flat current 
Universe is a new problem, which we label the quasi-flatness problem. 
We show how theories incorporating time-varying coupling constants
could drive the Universe to a late-time near-flat attractor.
A similar problem may be posed with regards to the cosmological constant
$\Lambda$, the quasi-lambda problem, and we exhibit a solution
to this problem as well.
\end{abstract}
\end{frontmatter}

\section{Introduction}

Recent investigations of the cosmological effects of introducing a
time-variation of the speed of light, $c$, into the gravitational field
equations have revealed a number of tantalising possibilities . If the speed
of light falls sufficiently rapidly over an interval of time then it is
possible to solve the standard horizon and flatness problems in a way that
differs from the inflationary universe, \cite{mof,vsl0,vsl1}. Moreover, unlike
in inflationary models, \cite{infl}, or pre big-bang scenarios, \cite{prebb}%
, a sufficiently fast fall-off in $c$ can also solve the cosmological
constant problem \cite{vsl0,vsl1,mof1}.

In this letter we display another appealing feature of these varying-$c$
models: they permit a solution of the \textit{quasi-flatness} and \textit{%
quasi-lambda }problems. Whereas a solution of the flatness
problem must provide for the flat $\Omega =1$ cosmology being a late-time
attractor for ever-expanding universes, a solution to the quasi-flatness
problem must offer an explanation for a late-time attractor with $\Omega \ $%
taking a value or order unity, between zero and one, for example $\Omega
_0\sim 0.2$. An inflationary universe model could only provide a resolution
of this problem by appealing to a particular (small) amount of inflation
that left the expansion short of asymptoting very close to flatness. The
motivation for seeking a model in which such an attractor with $\Omega <1$
arises naturally is the persistent trend of the observational evidence \cite
{coles}. It almost all points towards such a value in our Universe.

Similarly, whereas a solution of the lambda problem would require that the
contribution of the cosmological constant term, $\Lambda ,$ to the Friedmann
equation be negligible at late times in ever-expanding universes, \cite{lamb}%
, a solution of the quasi-lambda problem would explain why its effect is of
the same order as the density term in the Friedmann equation. The motivation
for this type of model is the recent evidence from supernovae suggesting
that the cosmological dynamics are best fitted by a model with $\Lambda \neq
0$, \cite{super}.

This type of concern has also stimulated new investigations in quantum
cosmology, \cite{open}, following earlier investigations of open
inflationary universes, \cite{turok}. It is possible for inflationary
universe models to provide a solution to the quasi-lambda problem if there
is a residual cosmological constant remaining at the end of inflation: the
dynamics then approach a model in which the curvature is negligibly small
and $\ \Omega _0=1-\Lambda /3H_0^2,$ where $H_0$ is the Hubble expansion
rate.

\section{The Quasi-flatness Problem}

The cosmological evolution of an isotropic and homogeneous universe
incorporating varying speed of light can be described by the standard
evolution equations for the Friedmann universe with $c=c(t):$

\begin{eqnarray}
\frac{\dot a^2}{a^2} &=&\frac{8\pi G(t)\rho }3-\frac{Kc^2(t)\ }{a^2},
\label{fr} \\
\ddot a &=&-\frac{4\pi G(t)}3(\rho +\frac{3p}{c^2(t)})a,  \label{acc}
\end{eqnarray}
where $p$ and $\rho $ are the density and pressure of the matter,
respectively, and $K$ is the metric curvature parameter. By differentiating (%
\ref{fr}) with respect to time and substituting in (\ref{acc}), we find the
generalised conservation equation \cite{vsl0} incorporating possible time
variations in $c(t)$,

\begin{equation}
\dot \rho +3\frac{\dot a}a(\rho +\frac p{c^2})=\frac{3Kc\dot c}{4\pi Ga^2}.
\label{cons}
\end{equation}

We shall assume that the matter obeys an equation of state of the form

\begin{equation}
p=(\gamma -1)\rho c^2(t),  \label{gam}
\end{equation}
where $\gamma $ is a constant. We shall assume that the universe undergoes a
sufficiently long period of evolution during which $c$ changes at a rate
proportional to the expansion of the universe:

\begin{equation}
c=c_0a^n,  \label{see}
\end{equation}
where $c_0>0$ and $n$ are constants, \cite{vsl1}.

Elsewhere, \cite{vsl0}, \cite{vsl2}, we have discussed the lagrangian
formulation of such a theory and its relationship, via transformations of
units, to a theory in which $c$ is constant (set equal to unity for 
example) and the electron charge varies.
This is just a manifestation of the fact that an invariant operational
meaning can only be attributed to the variation of dimensionless constants.
In this case, varying $e$ or $c$ would be different representations of a
theory displaying the observational consequences of varying fine structure
constant, in which there has been much recent observational interest \cite
{alpha}. An example of a theory with varying $e$ but constant $c$ has been
given by Bekenstein, \cite{bek}, but as shown in \cite{vsl2} the theory
leading to (\ref{fr})-(\ref{cons}) differs from Bekenstein's because he
postulates \textit{ab initio} that there be no effects on the gravitational
field equations.

The cosmological density parameter, $\Omega ,$ is defined in the usual way
as the density of the universe in units of the critical density, $\rho _c,$
which defines the $K=0$ solution of eq. (\ref{fr}). Thus,

\begin{equation}
\Omega \equiv \frac \rho {\rho _c}=\frac{8\pi G\rho }{3H^2}  \label{om1}
\end{equation}
and so, again using (\ref{fr}), we have 
\begin{equation}
{\frac \Omega {\Omega -1}}={\ }\frac{8\pi G\rho }{3Kc^2(t)a^{-2}}.\ 
\label{om2}
\end{equation}

We can solve eq. (\ref{cons}), using (\ref{gam}) and (\ref{see}), to obtain 
\cite{vsl1}

\begin{equation}
\rho =\frac B{a^{3\gamma }}+\frac{3Kc_0^2na^{2(n-1)}}{4\pi G(2n-2+3\gamma )},
\label{rho}
\end{equation}
where $B\geq 0$ is constant if $2n-2+3\gamma \neq 0$. When the speed of
light is constant ($n=0$) this reduces to the usual adiabatic density
evolution law $\rho \propto a^{-3\gamma }$. Substituting in (\ref{fr}) we
have

\begin{equation}
\frac{\dot a^2}{a^2}=\frac{8\pi GB}{3a^{3\gamma }}+\frac{%
Kc_0^2a^{2(n-1)}(2-3\gamma )}{(2n-2+3\gamma )}\   \label{fr1}
\end{equation}
and hence (\ref{om2}) reduces to

\begin{equation}
{\frac \Omega {\Omega -1}}={\ }\frac{8\pi G}{3Kc_0^2}\left[ Ba^{2-2n-3\gamma
}+\frac{3Kc_0^2n}{4\pi G(2n-2+3\gamma )}\right] .  \label{om3}
\end{equation}

These equations allows us to determine the possible cosmological behaviours
that can arise as  $n,$ the rate of variation of the speed of light, varies.
Thus, we see that if $2-2n-3\gamma >0$ then, as the universe expands, $%
a\rightarrow \infty $, the right-hand side of (\ref{om3}) diverges and we
have $\Omega \rightarrow 1$ and the expansion asymptotes to $a(t)\propto
t^{2/3\gamma }$ for $\gamma >0$; that is, we provide an explanation for 
\textit{flatness} at large $a$ or time, $t$, whenever, \cite{vsl0,vsl1},

\begin{equation}
n<\frac 12(2-3\gamma ).  \label{n1}
\end{equation}
By contrast, if $2-2n-3\gamma <0,$ then as $a\rightarrow \infty $, we have

\begin{equation}
{\frac \Omega {\Omega -1}}\rightarrow \frac{2n}{2n-2+3\gamma }  \label{om4}
\end{equation}
and hence we have a solution of the \textit{quasi-flatness} problem. We
predict that at large $a$

\begin{equation}
\Omega \rightarrow \frac{-2n}{3\gamma -2}  \label{om5}
\end{equation}
whenever

\begin{equation}
0>n>\frac 12(2-3\gamma )  \label{n2}
\end{equation}
where $n<0$ is required to ensure $\Omega >0$ when the equation of state
satisfies $3\gamma >2.$ From (\ref{fr1}) we see that at late times the scale
factor asymptotes to

\begin{equation}
a(t)\propto t^{\frac 1{1-n}}.  \label{a}
\end{equation}

For example, if the universe is radiation ($\gamma =4/3$) or dust ($\gamma =1
$) dominated then a solution of the quasi-flatness problem would require a
period of evolution during which $0>n(rad)>-1$ or $0>n(dust)>-\frac 12,$
leading to asymptotic expansion with $\Omega (rad)=-n$ and $\Omega
(dust)=-2n,$ respectively. Thus, asymptotic expansion with $\Omega _0$ less
than but of order unity is possible in these models. If the fluid which
dominates the expansion dynamics during the period when $c$ varies violates
the strong energy condition (as is required for inflation to occur), so $%
0<3\gamma <2,$ then a quasi-flat asymptote cannot arise because we also
require $n>0$ for $\Omega >0.$

The conditions required for the solution of the horizon and monopole
problems are identical to those for the flatness problem \cite{vsl0,vsl1}.

\section{The Quasi-lambda Problem}

Let us now consider the more stringent requirements on $c$ variation that
are required to resolve the problems associated with the possible existence
of a non-negligible cosmological constant term in the cosmological
equations. If we wish to incorporate a positive cosmological constant term, $%
\Lambda $, (which we shall assume to be constant) into a theory with varying
speed of light then we can define a vacuum stress obeying an equation of
state

\begin{equation}
p_\Lambda =-\rho _\Lambda c^2,  \label{vac}
\end{equation}
where

\begin{equation}
\rho _\Lambda =\frac{\Lambda c^2}{8\pi G}\geq 0.  \label{AL}
\end{equation}
Then, replacing $\rho $ by $\rho +\rho _\Lambda $ in (\ref{cons}), we have
the generalisation \cite{vsl0}

\begin{equation}
\dot \rho +3\frac{\dot a}a(\rho +\frac p{c^2})+\dot \rho _\Lambda =\frac{%
3Kc\dot c}{4\pi Ga^2}.  \label{cons2}
\end{equation}
The Friedmann equation is now

\begin{eqnarray}
\frac{\dot a^2}{a^2} &=&\frac{8\pi G\rho }3-\frac{Kc^2(t)\ }{a^2}+\frac{%
\Lambda c^2(t)}3.  \label{fr2} \\
&&\ \   \nonumber  \label{ac}
\end{eqnarray}
We define

\begin{equation}
\Omega \equiv \Omega _m+\Omega _\Lambda =\frac{8\pi G(\rho +\rho _\Lambda
)a^2}{3Kc^2}  \label{om6}
\end{equation}
where we distinguish the contributions from the matter ($m$) and the lambda
term ($\Lambda $) by the ratio

\begin{equation}
\frac{\Omega _m}{\Omega _\Lambda }\equiv \frac \rho {\rho _\Lambda }.
\label{om7}
\end{equation}

Again, we assume that $c\ $varies according to (\ref{see}), so eq. (\ref
{cons2}) integrates to give \cite{vsl1}

\begin{equation}
\rho =\frac B{a^{3\gamma }}+\frac{3Kc_0^2na^{2(n-1)}}{4\pi G(2n-2+3\gamma )}-%
\frac{\Lambda nc_0^2a^{2n}}{4\pi G(2n+3\gamma )}.  \label{rho2}
\end{equation}
Substituting in (\ref{fr2}) we have

\begin{equation}
\frac{\dot a^2}{a^2}=\frac{8\pi GB}{3a^{3\gamma }}+\frac{%
Kc_0^2a^{2(n-1)}(2-3\gamma )}{(2n-2+3\gamma )}+\frac{\Lambda \gamma
c_0^2a^{2n}}{(3\gamma +2n)}  \label{frnew}
\end{equation}
and

\begin{equation}
\frac{\Omega _m}{\Omega _\Lambda }=\frac{8\pi GBa^{-3\gamma -2n}}{\Lambda
c_0^2}+\frac{6Kna^{-2}}{\Lambda (2n-2+3\gamma )}-\frac{2n}{(2n+3\gamma )}.
\label{om8}
\end{equation}

Eq. (\ref{frnew}) allows us to determine what happens at large $a.$ We note
that the curvature term is always dominated by the $\Lambda $ term at
sufficiently large $a.$ Eq. (\ref{om8}) allows us to infer whether a
solution of the quasi-lambda problem is possible.

There are two cases to consider:

\subsection{Subcase 1: $n<-3\gamma /2$}

If $-3\gamma >2n$ then we see that the flatness and the lambda problems are
both solved as before since the $B$ term dominates the right-hand side of
eq. (\ref{frnew}) at large $a,$ with $a(t)\propto t^{2/3\gamma }$. In this
case there is no possible resolution of the quasi-flatness problem since,
from (\ref{om8}), we see that $\Omega _m/\Omega _\Lambda \rightarrow \infty .
$

\subsection{Subcase 2: $0>\ n>-3\gamma /2$}

$\ $In this case the $\Lambda $ term dominates the dynamics of eq. (\ref
{frnew}) yielding

\begin{equation}
\frac{\dot a^2}{a^2}\ \rightarrow \frac{\Lambda \gamma c_0^2a^{2n}}{(3\gamma
+2n)}.  \label{lamas}
\end{equation}
So, at large $t,$we have

\begin{equation}
a\propto t^{\frac{-1}n}.  \label{as1}
\end{equation}
However, for negative $n,$ now there is a solution to the quasi-lambda
problem (\textit{ie} why $\Omega _m$ and $\Omega _\Lambda $ are of similar
order at large $a$) since the ratio of the densities contributed by the
matter and lambda stresses approaches a constant positive value determined
by $n:$

\begin{equation}
\frac{\Omega _m}{\Omega _\Lambda }\ \rightarrow -\frac{2n}{2n+3\gamma }>0.
\label{om9}
\end{equation}
We can also express the asymptotic form of the scale factor as $t\rightarrow
\infty $ in the form

\begin{equation}
a(t)\propto t^\lambda  \label{a3}
\end{equation}
where

\begin{equation}
\lambda \equiv -\frac{3\gamma \Omega _m}{2(\Omega _\Lambda +\Omega _{m})}.
\label{a4}
\end{equation}

An interesting prediction of our model is that if $\Omega _\Lambda \neq 0$
then one must have $\Omega _m+\Omega _\Lambda =1$. Therefore if we are to
solve the quasi-lambda problem, we must have exact flatness. This is in
agreement with recent observational indications \cite{super}.

\section{Discussion}

We have shown that a cosmological theory in which the velocity of light
experiences a period of change can have a number of appealing consequences.
Elsewhere, we have shown that a suitable fall-off in the value of $c$ can
provide solutions to either or both of the flatness and cosmological
constant problems. The flatness problem is solved in a different way to its
resolution in inflationary universes: the curvature term is made to decrease
faster than the matter term in the Friedmann equation whereas in general
relativistic inflationary cosmologies the matter terms are made to fall off
more slowly than the curvature term by appeal to matter fields with $\rho
+3p<0$. Inflation does not resolve the lambda problem at all. Here we extend
those results by showing that solutions of the more difficult quasi-flatness
and quasi-lambda problems can also be found in such a scenario. For a simple
power-law change of $c=c_0a^n$ we have determined the range of values of $n$
which provide solutions of the flatness, quasi-flatness, lambda, and
quasi-lambda problems in the presence of matter with a perfect fluid
equation of state.

A number of open problems remain, to be investigated in a longer publication.
 Standard varying-$c$ scenarios solve the flatness problem due
to violations of energy conservation, as encoded in (\ref{cons}) and (\ref
{cons2}). These violations are only non-negligible when curvature
``attempts'' to dominate. Then, matter is produced if the Universe is open
(and therefore sub-critical), and destroyed if the Universe is closed (and
therefore super-critical). A changing $c$ introduces a self-regulating
mechanism which stabilizes the flat or critical model. The implications of
this process for the entropy of the Universe were also discussed in \cite
{vsl0}. In the scenarios we discuss in this Letter, on the other hand, the
energy source term is present even after the attractor is achieved. We have
therefore a scenario of ``permanent reheating'', which we will detail in 
a future publication. 

We will return the study of the general class of
solutions to this model elsewhere.
In particular we shall consider the effects of the
radiation to matter transition, and also the possibility of a changing $G$.
The latter results from the theories discussed in \cite{vsl1} which are a
generalization of the usual Brans-Dicke theories.

\section{Acknowledgments}

JM acknowledges financial support from the Royal Society and would like to
thank C. Santos for helpful comments. JDB is supported by a PPARC Senior
Fellowship.

\end{document}